\documentstyle{mn}

\newif\ifAMStwofonts
\ifoldfss

  \ifCUPmtlplainloaded \else
    \NewTextAlphabet{textbfit} {cmbxti10} {}
    \NewTextAlphabet{textbfss} {cmssbx10} {}
    \NewMathAlphabet{mathbfit} {cmbxti10} {} 
    \NewMathAlphabet{mathbfss} {cmssbx10} {} 
  \fi
  \ifAMStwofonts
    \ifCUPmtlplainloaded \else
      \NewSymbolFont{upmath} {eurm10}
      \NewSymbolFont{AMSa} {msam10}
      \NewMathSymbol{\upi}     {0}{upmath}{19}
      \NewMathSymbol{\umu}     {0}{upmath}{16}
      \NewMathSymbol{\upartial}{0}{upmath}{40}
      \NewMathSymbol{\leqslant}{3}{AMSa}{36}
      \NewMathSymbol{\geqslant}{3}{AMSa}{3E}

      \let\leq=\leqslant 
      \let\geq=\geqslant 
    \fi
  \fi
\fi 
\ifnfssone
  \newmathalphabet{\mathit}
  \addtoversion{normal}{\mathit}{cmr}{m}{it}
  \addtoversion{bold}{\mathit}{cmr}{bx}{it}

  \newmathalphabet{\mathbfit} 
  \addtoversion{normal}{\mathbfit}{cmr}{bx}{it}
  \addtoversion{bold}{\mathbfit}{cmr}{bx}{it}
  \newmathalphabet{\mathbfss} 
  \addtoversion{normal}{\mathbfss}{cmss}{bx}{n}
  \addtoversion{bold}{\mathbfss}{cmss}{bx}{n}
  \ifAMStwofonts
    \ifCUPmtlplainloaded \else
      \UseAMStwoboldmath
      \makeatletter
      \new@mathgroup\upmath@group
      \define@mathgroup\mv@normal\upmath@group{eur}{m}{n}
      \define@mathgroup\mv@bold\upmath@group{eur}{b}{n}
      \edef\UPM{\hexnumber\upmath@group}
      \new@mathgroup\amsa@group
      \define@mathgroup\mv@normal\amsa@group{msa}{m}{n}
      \define@mathgroup\mv@bold\amsa@group{msa}{m}{n}
      \edef\AMSa{\hexnumber\amsa@group}
      \makeatother
      \mathchardef\upi="0\UPM19
      \mathchardef\umu="0\UPM16
      \mathchardef\upartial="0\UPM40
      \mathchardef\leqslant="3\AMSa36
      \mathchardef\geqslant="3\AMSa3E

      \let\leq=\leqslant 
      \let\geq=\geqslant 
    \fi
  \fi
\fi 

\ifnfsstwo
  \DeclareMathAlphabet{\mathbfit}{OT1}{cmr}{bx}{it}
  \SetMathAlphabet\mathbfit{bold}{OT1}{cmr}{bx}{it}
  \DeclareMathAlphabet{\mathbfss}{OT1}{cmss}{bx}{n}
  \SetMathAlphabet\mathbfss{bold}{OT1}{cmss}{bx}{n}
  \ifAMStwofonts
    \ifCUPmtlplainloaded \else
      \DeclareSymbolFont{UPM}{U}{eur}{m}{n}
      \SetSymbolFont{UPM}{bold}{U}{eur}{b}{n}
      \DeclareSymbolFont{AMSa}{U}{msa}{m}{n}
      \DeclareMathSymbol{\upi}{0}{UPM}{"19}
      \DeclareMathSymbol{\umu}{0}{UPM}{"16}
      \DeclareMathSymbol{\upartial}{0}{UPM}{"40}
      \DeclareMathSymbol{\leqslant}{3}{AMSa}{"36}
      \DeclareMathSymbol{\geqslant}{3}{AMSa}{"3E}

      \let\leq=\leqslant 
      \let\geq=\geqslant 
    \fi
  \fi
\fi 
\ifCUPmtlplainloaded \else
  \ifAMStwofonts \else 
    \def\upi{\pi}
    \def\umu{\mu}
    \def\upartial{\partial}
  \fi
\fi
\title{On the Dust Extinction in High-$z$ Galaxies and the Case of Extremely 
Red Objects}
\author[A. Cimatti et al.]
       {Andrea Cimatti$^1$, Simone Bianchi$^{2,3}$,
	   Andrea Ferrara$^1$ and Carlo Giovanardi$^1$\\
	   $^1$ Osservatorio Astrofisico di Arcetri, Largo E. Fermi, 5, 50125
		Firenze, Italy\\
	   $^2$	Universit\`a di Firenze, Dipartimento di Astronomia e Scienza dello
	   Spazio, Largo E. Fermi, 5, 50125, Firenze, Italy\\
	   $^3$ Department of Physics and Astronomy, University of Wales Cardiff,
			  P.O. Box 913, Cardiff Wales, CF2 3YB, UK}
\date{}
\pagerange{\pageref{firstpage}--\pageref{lastpage}}
\pubyear{1994}
\begin{document}
\maketitle
\label{firstpage}
\begin{abstract}

We present the preliminary results of Monte Carlo simulations aimed 
to investigate the effects of realistic dust extinction (absorption +
scattering) on the colours of high-$z$ galaxies. In this paper, we
concentrate on the case of spheroidal galaxies, and we obtain attenuation 
curves in the range 0.1-1.2$\mu$m for different dust spatial distributions, 
and for a range of 
values of the dust optical depth, geometrical thickness and inclination of 
the dust disk. We find that the resultant curves are strongly dependent on 
the dust geometrical distribution and optical depth. A serendipitous finding is
that the strength of the 2200~\AA~ absorption feature depends not only on the
optical depth, but also on the dust geometrical distribution. As a first application, 
we test our results on two high-$z$ galaxies with extremely red colours (HR10 and
HR14) in order to infer clues on their ages, dust content and dust spatial
distribution. We confirm that HR10 must be a very dusty galaxy, and we suggest 
that its stellar component should be strongly embedded in the dust in order to
reproduce the observed extremely red colours.

\end{abstract}

\section{Introduction}

Internal dust might affect seriously the observed properties of distant 
objects, and observations have indeed found evidence of dust in high-$z$ 
objects. Several active galaxies with $z>$2 have been detected at  
submillimeter wavelengths, suggesting the presence of large amounts 
of dust (M$_{dust} \sim 10^{8-9}$ M$_{\odot}$; see Hughes 1996 for a recent 
review). The reddening of the background quasars and their metal abundances
suggest the presence of dust in damped Ly$\alpha$ absorption 
systems (Pettini et al. 1994; Pei \& Fall 1996). Also, the UV polarization
properties of high-$z$ radio galaxies can be explained in term of dust
scattering, suggesting a significant amount of dust in their ISM (Cimatti 
1996 and references therein). A substantial amount of dust is also expected 
in evolutionary models of spheroidal galaxies at high-$z$ (Franceschini et al. 1994;
Mazzei \& De Zotti 1996 and references therein). 

Although dust is likely to be present in most high-$z$ systems, 
no information is available neither on its spatial distribution,
nor about how {\it realistic} dust extinction can affect our view of 
distant galaxies. In fact, extinction is usually treated in a 
simplistic way, neglecting the contribution of scattering, and assuming naive 
spatial distributions (e.g. uniform foreground screens or infinite slabs). Only 
recently, 
more realistic models have been developed (Kylafis \& Bahcall 1987; Bruzual et al. 
1988; Witt et al. 1992; Byun et al. 1994; Wise \& Silva 1996; Bianchi, Ferrara 
\& Giovanardi 1996, hereafter BFG). Although these models have shown that dust 
scattering plays an important role by reducing the effects of dust 
absorption, no systematic studies of the effects of the extinction on the 
observed colours of high-$z$ galaxies have been performed. Understanding 
these effects is crucial in cosmology and galaxy evolution studies. 

For instance, a relevant problem arises in the age estimates of high-$z$ galaxies. 
When deep spectroscopy is not available, no information on the stellar 
continuum and absorption features are obtainable, and the age estimates 
are based solely on the fitting of the broad-band photometric Spectral Energy
Distributions (SEDs) with synthetic stellar population spectra. 
In a simplistic scenario where the extinction is entirely due to absorption
(the screen model), it is well known that the colours of 
a reddened young galaxy can mimic those of an unreddened old galaxy, 
producing a degeneracy of the age and dust extinction. However, what happens
to the integrated colours of a galaxy when dust extinction is treated more 
realistically is poorly known. 

Witt et al. (1992) suggested that the effects 
of realistic dust extinction on the SEDs of high-$z$ galaxies are not strong. 
However, their claim was based on a single object (the radio galaxy B2 0902+34
at $z\sim3.4$), and generalized to the whole population of high-$z$ galaxies.
In addition, the continuum SED of B2 0902+34
turned out to be completely different from that used by Witt et al. (1992)
because of a strong contamination of the $K$-band flux by the redshifted
[OIII]$\lambda\lambda$4959+5007 (Eisenhardt \& Dickinson 1992). 

Instead, recent studies (Franceschini et al. 1994) have emphasized
the role of dust in the early stages of evolution of galaxies.
The present optical surveys for high-$z$ galaxies are designed 
to select objects with strong emission lines, such as Ly$\alpha$,
(see Pritchet 1994 for a review; Thompson et al. 1995), or galaxies 
with a flat continuum spectrum and with a sharp Lyman-break (Steidel et al. 1996).
However, because of those selection criteria, these surveys would miss the putative 
population of high-$z$ galaxies obscured by dust extinction. It is notable that the 
strongest evolutionary effects due to dust extinction are not expected for the galactic 
disks, or in disk dominated systems (Mazzei et al. 1992), but rather in early-type 
galaxies which, under appropriate circumstances, might experience a prolonged opaque 
phase (Mazzei \& De Zotti 1996).  The whole issue of the effects of dust on the SEDs 
of high-$z$ galaxies still remains an important, open question, especially in view of 
the future ISO and sub-mm data.

Extinction effects become particularly relevant in the case of high-$z$ galaxies 
with very red colours, also called extremely red objects (EROs) (McCarthy et al. 
1992; Hu \& Ridgway 1994). In fact, these galaxies may be distant and old ellipticals, 
and could provide crucial clues on the first epoch of galaxy formation, $H_0$ and $q_0$,
once their ages are derived accurately. However, they could also be very dusty high-$z$
galaxies whose intrinsic colours are strongly reddened by dust extinction.
The main problem is that the degeneracy 
of the age and the dust extinction becomes maximized in this class of objects 
because of the similarity between the colours of a genuinely old galaxy and those 
induced by foreground screen reddening. 
The main questions are then : how much {\it realistic} dust extinction can mimic 
the colours of an old galaxy ? If these galaxies are dusty, is it possible to 
learn something about the spatial distribution of the dust by modeling their SEDs ?

Motivated by the general lack of information on the effects of dust extinction on the
SEDs of high-$z$ galaxies, we have started an extensive study aimed at 
investigating 
the relevance of these effects. In this paper, we focus on the case of 
spheroidal galaxies, present the first results, 
discuss an application to the case of the extremely red galaxies, and show a
serendipitous result about the strength of the 2200~\AA~ dust absorption feature in 
external galaxies. More general results and applications will be presented in a 
forthcoming paper.

\section{Monte Carlo Simulations}

As a first attempt to study the effects of dust extinction, we
investigate the case of spheroidal galaxies. The stellar density 
profile $\rho_{star}(r)$ is modelled as a Jaffe bulge (see BFG), 
which reproduces the surface brightness $r^{1/4}$ profile characteristic 
of elliptical and bulge systems (de Vaucouleurs 1959). We adopt an 
effective radius $r_e$=4~kpc, representative of nearby elliptical and 
bulge systems (Binney \& Tremaine 1987), and the distribution is truncated 
at a radius $r_{max}=5 r_e=20$~kpc (see BFG for more details). 

Dust extinction has been treated by using Monte Carlo simulations of realistic 
radiative transfer in dusty galaxies (i.e. considering the extinction as the 
combination of absorption and scattering). The details of the model are not
repeated here and can be found in BFG. However, it is important to recall here
that we assume Galactic dust: in our model the grains are supposed to be spherical 
and to have a size distribution given by the MRN model (Mathis, Rumpl \& Nordsiek
1977), $n(a)\propto a^{-3.5}$, where $a$ is the grain radius. We consider three 
materials: astronomical silicates, $\parallel$ graphite, and $\perp$ graphite.
The numerical silicates/graphite ratio is 1:1.12, with 1/3 of the graphite having 
optical properties measured parallel ($\parallel$), and 2/3 perpendicular ($\perp$) 
to the c-axis. The lower and upper limits of the distribution are $a_{-}=0.005$~
$\mu$m and $a_{+}=0.25$~$\mu$m, irrespectively of the material. Very small grains 
and PAH have not been included due to the large uncertainties both in their size 
distribution and optical constants. The dielectric constants adopted are the 
ones given by Draine \& Lee (1984), recently extended in the far UV and X-rays 
by Martin \& Rouleau (1991). All the relevant optical properties (absorption and 
scattering cross section, albedo) have been calculated using Mie formulae.

We tested three different spatial distributions for the dust within the
galaxies : {\bf (A)} a homogeneous disk with constant dust density 
$\rho_{dust}$  and with radius $r_d=r_{max}=20$~kpc equal to that of the 
stellar distribution. We considered three different values of the 
geometrical half-thickness of the disk, $z_d=70, 250, 500$~pc, and three different 
inclinations to the line of sight, $i=10^\circ, 45^\circ, 90^\circ$ 
(edge-on); {\bf (B)} a homogeneous spherical distribution with 
radius $r_d=r_{max}=20$~kpc equal to that of the stellar bulge, where 
the dust is intermixed with the stars and has constant density 
($\rho_{dust}(r)=$constant); {\bf (C)} a spherical distribution where the 
dust is intermixed with the stars, but its density follows the same 
radial distribution of the stellar component (i.e. $\rho_{dust}(r)=
\rho_{star}(r)$).

Model A can be considered representative of a spheroidal galaxy 
where the dust is located in a disk, similarly to what is actually
observed in a number of nearby ellipticals (Goudfroij 1996, and references 
therein). On the other hand, in models B and C, the dust is 
interspersed with the stars. In particular, model B, although simplistic, 
can mimic a very dusty spheroidal galaxy where the stars are strongly
embedded in the dust, and it is somehow similar to the classic foreground
screen model, i.e. where the dust has a large geometrical covering factor,
and where dust absorption dominates. 
Finally, model C can be considered representative of a spheroidal galaxy 
where the dust is distributed more realistically, following the same 
gravitational potential as the stellar component. We are aware that 
high-$z$ galaxies may have different distributions of dust, and that
models A, B and C have simplistic geometrical configurations. However, 
these models allow us to test on the observed SEDs of high-$z$ galaxies 
a broad variety of circumstances, from very dusty cases (model B) to 
cases where the dust is located only in a disk (model A), and have the
capability to treat dust extinction realistically, i.e. taking into account 
both absorption and scattering. 

In addition to the above geometrical parameters, each model is characterized 
by the value of the optical depth in the V-band in the rest frame of the
galaxy, $\tau_V$, along a line of sight passing through the centre of the
galaxy for models A and B (perpendicular to the plane, in the disk case). 
In the case of model C, due to the central cusp of the $r^{1/4}$ law,
it is preferable to define $\tau_V$ for a line of sight at projected
distance $1.16~r_e$ from the centre (see BFG). The simulations were
carried out for a range of $\tau_V$ from 0.1 to 10, and for 28 
wavelengths in the range 0.1--1.2$\mu$m.

\section{Results}

\subsection{Attenuation curves}

From the Monte Carlo simulations we obtain the attenuation curves
${\cal A}_\lambda$
(defined as the ratio of the observed, dust extincted, total intensity, 
$I_{obs}(\lambda)$ to the intrinsic, unextincted, one of the galaxy, $I_0
(\lambda)$) as a function of wavelength for different values of $\tau_V$, 
$z_d$, and $i$ (see Fig. 1). 

As expected, in the case of the disk (model A) with $i=90^\circ$, we find 
an almost flat ${\cal A}_\lambda$ for any value of $\tau_V$ and $z_d$ (i.e. no 
effects on the colours). For inclinations $i<90^\circ$, we find instead 
a strong dependence of ${\cal A}_\lambda$ on $\tau_V$, 
whereas the depencence on $z_d$, and $i$ is much weaker. A 
representative case, $z_d$=250 pc and $i=45^\circ$, is shown in Fig. 1:
the shape of the curve changes 
significantly from $\tau_V \sim$0.1-1.0 to $\tau_V >$1.0, and the depth 
of the 2200~\AA~ feature varies considerably, reaching a maximum for $\tau_V 
\sim$1, becoming weaker for larger $\tau_V$, and almost disappearing 
for $\tau_V>4$. Regarding the case of the disk, it is important to
note that we found that an exponential distribution of the dust
within the disk does not produce any relevant difference with respect to
the case of a disk with a constant dust density.

For the bulge with constant dust density (model B), also shown in Fig. 1, the 
slope steepens 
considerably for $\tau_V>1$, and the depth of the 2200~\AA~ feature 
shows remarkable variations as a function of $\tau_V$. Moreover, at a
given $\tau_V$, model B produces attenuation curves much redder than
in model C.

In summary, for a given $\tau_V$ there are appreciable differences between 
the three models in terms of the slope of the ${\cal A}_\lambda$ and strength 
of the 2200~\AA~ feature. A detailed analysis of these results will be 
presented in a forthcoming paper.

\begin{figure}
\vspace{11cm}
\caption{A representative set of attenuation curves obtained for the three
models : model A (solid), B (short-dashed), and C (long-dashed). The curves are
shown for $\tau_V$=0.1,0.5,1.0,2.0,4.0,10.0, with $\tau_V$ increasing from the top to
the bottom of the figure.}
\end{figure}

\subsection{The 2200~\AA~ feature and its absence in external galaxies}

The absence or weakness of the 2200~\AA~ absorption feature has been 
recently debated extensively  (Calzetti et al. 1994 and references therein;
see also Calzetti et al. 1995). The problem arises when the 2200~\AA~ feature 
is expected, but not observed in galaxies which are known to be dusty from 
their emission line ratios and UV--optical continuum properties.
The absence of the feature has been usually ascribed either to a different
dust chemical composition or to scattering effects which compensate for
the dust absorption.

Our study shows that a peculiar chemical composition of the dust grains,
or the role of scattering may not be the only causes to explain the lack 
or weakness of the 2200~\AA~ feature in external galaxies. Figure 1 clearly
shows how the strength (measured by its equivalent width) of the 2200~\AA~ 
feature is a strong function of {\it both} the optical depth {\it and} of 
the geometry. The feature shows strong variations as a function of the optical 
depth within each of the models A, B and C. For instance, in model A (disk case) 
its equivalent width varies remarkably depending on $\tau_V$, reaching a maximum 
value of $\sim 
500$~\AA~ at $\tau_V\sim 1$ and decreasing to $\sim 65$~\AA~ at $\tau_V\sim 6$. 
In addition, strong variations are also evident, for a fixed $\tau_V$, as a 
function of the geometry of the dust extinction. For instance, for $\tau_V=1$,
the 2200~\AA~ feature displays large differences among the models A, B and C.
In this regard, it is important to note that, at least for the disk 
distribution (model A), the feature is strongly weakened or suppressed because 
the emission from opaque galaxies is mostly contributed by stars outside the 
dust disk and hence unaffected by absorption. 

The strong variations of the feature as a function of the geometric 
configuration suggest that the compensating effect of scattering may not be 
the only way to reduce the strength of the feature, and that the geometry 
may have a dominant influence. It is important to recall here that
the main contribution to the bump comes from small ($a\leq 0.1 \mu$m) graphite grains; 
for the dust composition here adopted, the weighted average radius at
this wavelength is in fact $\sim 0.05~\mu$m. Correspondingly, the albedo is
rather low, $\langle \tilde\omega \rangle \simeq 0.2$, and the scattering
phase function rather isotropic,  $\langle g \rangle \simeq 0.2$.
Thus, it seems unlikely that scattering is the main, or the only, origin 
of the weakness of the 2200~\AA~ feature compared to the pure absorption case.
However, a more detailed analysis is required to answer this question unambiguously.

Although our results suggest that the strength of the 2200~\AA~ feature may be 
a strong function of the geometrical effects, they do not rule out the possibility 
that external galaxies may have chemical grain properties different from those 
of our Galaxy. In fact, it is important to recall that the 2200~\AA~ feature is 
weak in the LMC, and absent in the SMC (Prevot et al. 1984). The issue of the 
2200~\AA~ feature will be investigated in details in our forthcoming paper.

\section{Application to extremely red galaxies}

As a first application, we investigate the effects of dust extinction on the 
SEDs of two extremely red galaxies at high-$z$. 
Extremely red objects, EROs, (i.e. $R-K > 6$) are found both in random
sky fields or in AGNs fields (Elston et al. 1988; McCarthy, Persson \& West
1992; Hu \& Ridgway 1994). Since their colours may be strongly affected by dust 
extinction, these galaxies represent an interesting application of our models, 
and suitable objects to investigate at what level the SEDs of dusty younger galaxies 
can mimic those of old ellipticals once we treat dust extinction in a realistic way, 
and to constrain the spatial distribution of the dust. 

We have selected from the literature two extremely red galaxies, HR10 and HR14, 
for which the best optical-IR photometric SEDs are available (Hu \& Ridgway 1994;
Graham \& Dey 1996). Our study is more concentrated on HR10, the only 
ERO with spectroscopically determined redshift.

We attempted to fit the SEDs of HR10 and HR14 with synthetic stellar population 
spectra, and applying the attenuation curves derived by our models in order to
take into account the possible dust extinction occurring in these galaxies. In 
our experiments we made use of the synthetic, unextincted, stellar population spectra 
of Bruzual \& Charlot (1993), with instantaneous burst of star formation, solar metallicity, 
and Salpeter IMF. 

Figure 2 shows our results for HR10, a galaxy at $z$=1.44 with a colour index 
$I-K^{'}$=6.5. Since its quoted $B$-band flux is not statistically significant 
($\sim2.2\sigma$ level), we adopt a $3\sigma$ upper limit. We also correct for 
the $\sim$20\% H-band flux contamination from the H$\alpha$ line. Graham \& Dey 
(1996), adopting a foreground screen dust absorption model, found that the SED 
of HR10 cannot be reproduced without dust extinction. Our independent analysis 
confirms that dust extinction must play an important role in HR10, 
and even very old (i.e. very red) stellar populations (up to 7 Gyr) cannot reproduce its 
SED without extinction. However our modeling allows us to go a step further, 
and to derive clues on the spatial distribution of the dust in HR10. In fact,
we find that model A (dust disk) cannot fit the data because the observed colours 
are always too red compared to those produced by this model for any $\tau_{V}$, 
age, $z_{d}$, and $i$. Model C allows a better fit to the SED, but the reddening 
is still not sufficient to match the $I-K^{'}$ colour. We find that only model
B gives an acceptable fit to the SED (age $\sim$ 1 Gyr and $\tau_{V}$=6). 
The SED can be fit also by a 1.0 Gyr old population reddened by a dust
foreground screen with $\tau_V$=1.54 (or $E_{B-V}$=0.55, adopting the Galactic 
relation $A_V \sim 3 E_{B-V}$). If we take these results at face 
value, we infer that the dust in HR10 is not distributed in a disk, and may be 
interspersed with the stars in a way that dust absorption dominates, resembling
the effects of a simple screen model, i.e. the stars may be strongly
embedded in the dust. This also shows how such modeling may provide 
new clues on the internal dust distribution in unresolved, distant galaxies.
The H$\alpha$ emission line properties observed by Graham \& Dey (1996) suggests 
that HR10 may be a very dusty galaxy with star formation and/or AGN activity. 
Adopting the model B with $\tau_V=6$ and a 1 Gyr old stellar population, we
estimate that the rest-frame dereddened K-band absolute magnitude is
$M_{K_{rest}}$=-27.5 ($H_0=50$ km s$^{-1}$ Mpc$^{-1}$, $q_0$=0.5), about
9 times brighter than $M_{K}^{*}$ of the luminosity function of the local field 
galaxies (Mobasher, Sharples \& Ellis 1993). 
%

\begin{figure}
\vspace{11cm}
\caption{The rest-frame UV--optical SED of HR10. Flux data points are from Graham 
\& Dey (1996). The curves are synthetic stellar population spectra extincted
by dust. The observed SED cannot be fit without extinction, but only model B is 
capable to produce enough reddening to match the colours of the galaxy.} 
\end{figure}

Figure 3 illustrates the results relative to HR14, a galaxy with $I-K^{'}$=6.2
(Hu \& Ridgway 1994). The case of this galaxy is more ambiguous because its
spectroscopic redshift is not known. Hu \& 
Ridgway (1994) estimated a photometric redshift $z_{phot}$=2.3, with a $\pm3\sigma$ 
acceptable fit range 1.8$<z<$3.0. Although its redshift is uncertain, we have
decided to investigate the case of this galaxy in three different cases :
$z$=1.8, 2.3 and 3.0, which represent the acceptable range of photometric
redshifts. For $z$=1.8, we find that its SED can be reproduced successfully 
without any dust extinction and with very old stellar populations ($\geq$4 Gyr). 
However, if dust extinction is taken into account, the SED can be fit within the 
errors by models A, B and C, and with a wide range of stellar population ages 
and $\tau_{V}$ (for instance, model A: $\tau_V\sim$ 3.0, age 1.6 Gyr; model B: 
$\tau_V\sim$ 2.0, age 1.6 Gyr; model C : $\tau_V\sim$ 2.0, age 2 Gyr). For $z$=2.3, 
the results are shown in Figure 3. The main result does not differ much from the 
case of $z$=1.8 : shifting HR14 to $z$=2.3 has the effect of requiring younger 
ages for all cases compared to $z$=1.8. It is also interesting to note that the 
foreground screen (absorption only) requires $\tau_V \sim 0.84$. For $z$=3.0, we 
find results almost identical to those of $z$=2.3, except for the $\tau_V$=0 case, 
where the SED is best fit with a slightly younger age (2.5 Gyr instead of 3.0 Gyr).

As a general result of this study, we note how realistic dust extinction
(i.e. absorption + scattering) requires larger dust optical depths
compared to the case of the pure absorption screen model (see also Witt et 
al. 1992). This is evident for example in Figure 1, where in order to
produce the same amount of reddening, it is required $E_{B-V}$=0.55 (i.e. 
$\tau_V \sim 1.5$), and $\tau_V \sim 6$ for the screen model and model B 
respectively.

\begin{figure}
\vspace{11cm}
\caption{The observed SED of HR14. Flux data points are from Hu \& Ridgway (1994),
and the case of $z$=2.3 is shown. The curves are synthetic stellar population 
spectra which provide acceptable fits to the observed SED. The SED of HR14 can
be also fit with an unextincted old stellar population spectrum.} 
\end{figure}

\subsection{Can HR10 be a dusty quasar ?}

Hu \& Ridgway (1994) found that the surface density of EROs 
is comparable to that of quasars. This result raises the question 
of whether EROs are dusty quasars and AGN missed by optical surveys. We
concentrate this discussion on HR10, the only ERO with a useful set of data
to attempt to answer this question.

First of all, it is necessary to check whether the photometry of HR10 may be
contaminated by strong emission lines, such as [OIII]$\lambda\lambda$4959+5007
in the J-band. Let us suppose that HR10 is a heavily obscured quasar. Adopting a 
typical ratio [OIII]/H$\alpha\leq$0.25 observed in 
Seyfert1/quasars (Netzer 1990), and using the observed H$\alpha$ flux (Graham \& 
Dey 1996), we find that the J-band flux contamination is small ($<$8\%).
We attempted to fit the SED of HR10 with a reddened quasar spectrum and a 
foreground dust absorption screen model. For this purpose, a point-like source
(the quasar nucleus) embedded in a dusty medium, the screen model is a good
approximation for the dust extinction. For the quasar spectrum, we used the
average spectrum of radio-quiet quasars of Cristiani \& Vio (1990) ($F_{\nu}
\propto \nu^{-0.7}$) because HR10 is a radio-quiet object. We find that the SED 
of HR10 cannot be reproduced with a reddened quasar spectrum for a wide range of 
$E_{B-V}$. We also tried to use flatter ($F_{\nu}\propto \nu^{-0.2}$) and
steeper ($F_{\nu}\propto \nu^{-1.2}$) quasar spectra in order to take into
account the dispersion of the slope of the average quasar spectrum, but
we always found that the SED cannot be reproduced by a reddened quasar
spectrum.

Therefore, although the H$\alpha$+[NII] line luminosity (3$\times10^{42} h_{100}
^{-2}$ erg s$^{-1}$; Graham \& Dey 1996) may be comparable to that of Seyfert 1
and quasars, our analysis of the SED suggests that the spectrum of HR10 is 
not dominated by AGN radiation. However, the large FWHM of the line (7000$\pm$ 
3000 km s$^{-1}$) seems to favour the presence of an AGN, or to indicate that part of 
the radiation comes from an active nucleus with broad lines. Deeper observations 
are needed to establish the value of the FWHM, which is only marginally significant.

\section{Summary and Conclusions}

We have presented the preliminary results of a set of simulations aimed to 
study the effects
of realistic dust extinction on the colours of high-$z$ galaxies. As a first
attempt, we have considered three different spatial distributions of the dust
in spheroidal galaxies, adopting Galactic dust grain properties. 
A serendipitous finding is that the 2200~\AA~ feature is generally much
weaker than in the foreground absorption screen model. The weakening of the feature is
particularly evident in the case of a disk of dust in a spheroidal galaxy, where
the feature can be almost absent. We find that 
the equivalent width of the 2200~\AA~ feature depends on the dust distribution, geometry 
and optical depth. We suggest that this effect is mostly a geometrical one, and 
that it might be one explanation of the failure of the detection of the 
2200~\AA~
feature in external dusty galaxies. However, other causes of the non-detection
of 2200~\AA~ feature, such as different chemical composition of the dust grains,
cannot be ruled out by the present results.

As a first application of our extinction curves, we consider the case of 
two extremely red galaxies at high redshift, HR10 and HR14, in order to
constrain the ages of their stellar populations, the amount of dust,
and the dust geometrical distribution.
Our results confirm that HR10 must be a very dusty galaxy, and allow us to
constrain the spatial distribution of the dust in this object.
However, we find that a strong degeneracy between the age of the stellar
population and the dust extinction is present in both galaxies.
In the case of HR10, we find that its colours can be reproduced by a 1 Gyr old
stellar population and $\tau_V\sim$6 (model B). Its dereddened rest-frame 
K-band absolute magnitude, $M_{K_{rest}}$=-27.5, is just beyond the highest 
luminosities of the local field galaxies. 

For HR14, although the redshift is poorly 
constrained, we find that its SED can be reproduced without extinction and with 
an old stellar population (2.5-4.0 Gyr) over the wide range 1.8$<z<$3.0. This
makes HR14 a good candidate of a mature galaxy at $z>$1.8. However, 
if dust extinction is present, the age can be lowered down to 0.8 Gyr for $\tau_V
\sim$2-3 and for a broad range of dust spatial distributions. It is important to 
note that the main results on HR14 are not strongly dependent on $z$, if 1.8$<z<$3.0. 
If HR14 would turn out to be at the same redshift of HR10, we would find results 
very similar to those obtained for HR10 because of their similar $I-K'$ colours.
We also explore the possibility that HR10 is a heavily obscured quasar, but we
find that its SED cannot be reproduced by a reddened quasar spectrum. However,
the present data cannot exclude that part of the radiation
in HR10 is due to an AGN.

Since realistic dust extinction requires larger optical depths 
compared to screen models (see also Witt et al. 1992), it is important 
to estimate the total amount of dust required by our models.
For standard values of the average grain size ($a\sim 0.1\mu$m) and density 
($\delta \sim 2$~g~cm$^{-3}$) the total dust mass in these galaxies is in the 
range $M_{dust}\sim \delta a R_{max}^2\tau_V=10^{8-9}$ M$_{\odot}$ for $\tau_V=
1-10$, respectively. In particular, taking into full account the dust composition, 
our models have $M_{dust}=k \, \tau_V \, 10^8$ M$_{\odot}$, 
where $k$ depends on the model geometry ($~k_A=1.21, ~k_B=0.81, ~k_C=0.35$).
These masses are comparable with those derived for several active
high-$z$ galaxies by submillimetric observations (Hughes 1996).

The existence of a population of extremely red galaxies, whose colours
can be explained only in terms of strong dust extinction, may be 
relevant to our understanding of the galaxy evolution. In fact, 
a dusty phase is predicted to occur during the early evolution of
spheroidal galaxies (Franceschini et al. 1994; Mazzei \& De Zotti 1996 and
references therein; Zepf \& Silk 1996). If this is correct, a substantial
population of dusty galaxies is expected to be present in the early
universe. These galaxies would be missed by the current optical 
selection criteria used to find high-$z$ galaxies (see
Pritchet 1994,PASP,106,1052; Steidel et al. 1996). However they would be selected
by deep surveys in the sub-mm spectral range, where the redshifted
dust thermal emission should peak (Blain \& Longair 1993). It is tempting to 
speculate whether the population of very red galaxies recently found
by deep optical and near-IR imaging (Hu \& Ridgway 1994; Knopp \&
Chambers 1997; Hogg et al. 1997) may be somehow related to this putative
family of high-$z$ galaxies hidden by dust obscuration. Future
multiwavelength observations of the extremely red galaxies will
provide new clues on their nature and role in the galaxy evolution picture.
 
\section*{Acknowledgments}
We are grateful to Paul Goudfroij for useful discussions, to
Ester Hu for providing the photometry of HR10 and HR14, and to
the anonymous referee for the useful suggestions.

\label{lastpage}

\begin{thebibliography}{99}
\bibitem{bi96} Bianchi, S., Ferrara, A. \& Giovanardi, C. 1996, ApJ, 465,
127 (BFG)
\bibitem{binn} Binney, J.  \& Tremaine S., 1987 in Galactic Dynamics,
(Princeton: UnivPress)
\bibitem{blain} Blain, A.W, \& Longair, M.S. 1993, MNRAS, 264, 509
%
\bibitem{br93} Bruzual, G. \&  Charlot, S. 1993, ApJ, 405, 538
\bibitem{br88} Bruzual, G., Magris G. \& Calvet, N. 1988, ApJ, 333, 673
\bibitem{bf94} Byun, Y. I., Freeman, K. C., \& Kylafis, N. D. 1994, ApJ, 432, 114
\bibitem{ca94} Calzetti, D., Kinney, A.L. \& Storchi-Bergmann, T. 1994, ApJ, 
429, 582
\bibitem{ca95} Calzetti, D., Bohlin, R.C., Gordon, K.D., Witt, A.N., Bianchi,
L. 1995, ApJ, 446, L97 
\bibitem{ci96} Cimatti, A. 1996, in "New Extragalactic Perspectives
in the New South Africa -- Changing Perceptions of the Morphology,
Dust Content and Dust-Gas Ratios in Galaxies", eds. D.L. Block \& J.M.
Greenberg, Kluwer Academic Publishers, p. 493 
\bibitem{cv90} Cristiani, S. \& Vio, R. 1990, A\&A, 227, 385
\bibitem{dV59} de Vaucouleurs, G., 1959, In "Handbuch der Physick", Vol. 53,
(Springer-Verlag: Berlin)
\bibitem{dra} Draine, B.T., \& Lee, H.M. 1984, ApJ, 285, 89
\bibitem{ei92} Eisenhardt, P., \& Dickinson, M. 1992, ApJ, 399, L47
\bibitem{e89} Elston, R., Rieke, G.H., \& Rieke, M.J. 1988, ApJ, 331, L77
\bibitem{fra} Franceschini, A., Mazzei, P., Danese, L., \& De Zotti, G. 1994, ApJ,
427, 140
\bibitem{go96} Goudfroij, P. 1996, in "New Extragalactic Perspectives
in the New South Africa -- Changing Perceptions of the Morphology,
Dust Content and Dust-Gas Ratios in Galaxies", eds. D.L. Block \& J.M.
Greenberg, Kluwer Academic Publishers, p. 400 
\bibitem{gh96} Graham, J.R. \& Dey, A. 1996, ApJ, 471, 720 
\bibitem{Hogg} Hogg, D.W., Neugebauer, G., Armus, L., Matthews, K., Pahre, M.A.,
Soifer, B.T., \& Weinberger, A.J. 1997, AJ, 113, 474
\bibitem{hu94} Hu, E.M. \& Ridgway, S.E. 1994, AJ, 107, 1303
\bibitem{hu96} Hughes, D.H.A 1996, in "Cold Gas at High Redshift",
eds. M. Bremer \& P. van der Werf, H. R\"ottgering \& C. Carilli, 
Kluwer Academic Publishers, p. 311
\bibitem{ky87} Kylafis, N. D., \& Bahcall, J. M. 1987, ApJ, 317, 637
\bibitem{knopp} Knopp, G.P., \& Chambers, K.C. 1997, ApJS, 109, 367
\bibitem{mar} Martin, P.G., \& Rouleau, F. 1991, in Extreme Ultraviolet Astronomy,
ed. R.F. Malina \& S. Bowyer (Elmsford, NY: Pergamon)
\bibitem{mrn} Mathis, J.S., Rumpl, W., Nordsiek, K.H. 1977, ApJ, 217, 425
\bibitem{ma96} Mazzei, P. \& De Zotti, G. 1996, MNRAS, 279, 535
\bibitem{ma92} Mazzei, P., Xu, C., \& De Zotti, G. 1992, A\&A, 256, 45
\bibitem{mc92} McCarthy, P.J., Persson S.E., \& West S. 1992, ApJ, 386, 52
\bibitem{m93} Mobasher B., Sharples R.M., \& Ellis R.S. 1993, MNRAS, 560, 574
\bibitem{n90} Netzer H. 1990, in {\it Active Galactic Nuclei}, Saas-Fee Advanced
Course 20, Springer-Verlag, p. 57
\bibitem{pe96} Pei, Y. C., \& Fall, S. M.  1996, ApJ, 454, 69
\bibitem{pe94} Pettini, M., Smith, L. J., Hunstead, R. W. \& King, 
D.L. 1994, ApJ, 426, 79
\bibitem{p84} Prevot, M.L., Lequeux, J., Prevot, L., Maurice, E.,
\& Rocca-Volmerange, B. 1984, A\&A, 132, 389
\bibitem{pr} Pritchet, S. 1994, PASP, 106, 1052
\bibitem{ste} Steidel C.C., Giavalisco M., Pettini M., Dickinson M., \& Adelberger
K.  1996, ApJ, 462, L17
\bibitem{th95} Thompson, D., Djorgowski, S., \& Trauger, J. 1995, AJ, 110, 963
\bibitem{wi96} Wise, M. W. \& Silva, D. R. 1996, ApJ, 461, 155
\bibitem{wi92} Witt, A.N., Thronson, H.A., \& Capuano, J.M. 1992, ApJ, 393, 611 
\bibitem{z} Zepf, S.E., \& Silk, J. 1996, ApJ, 466, 114
\end{thebibliography}
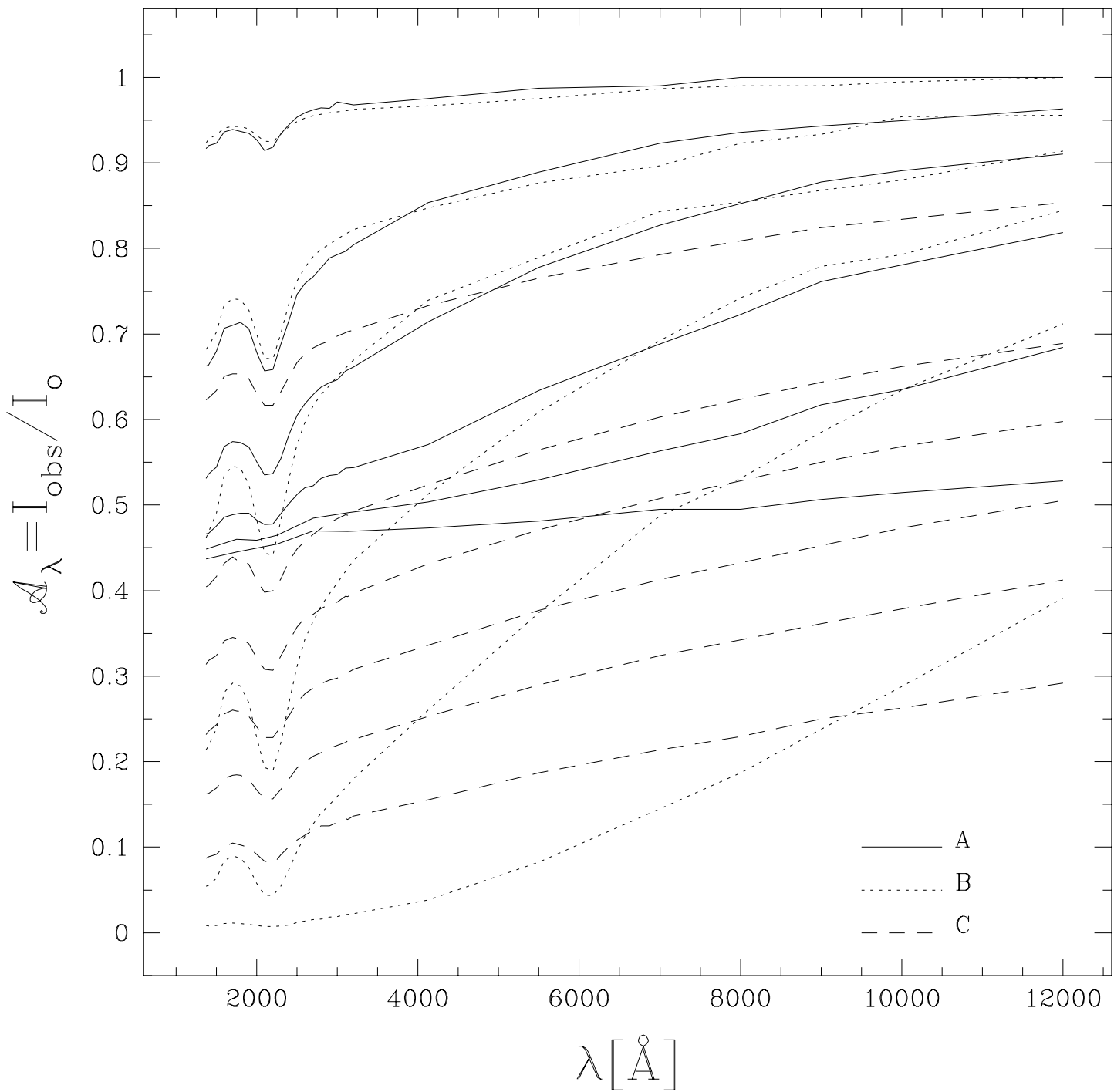
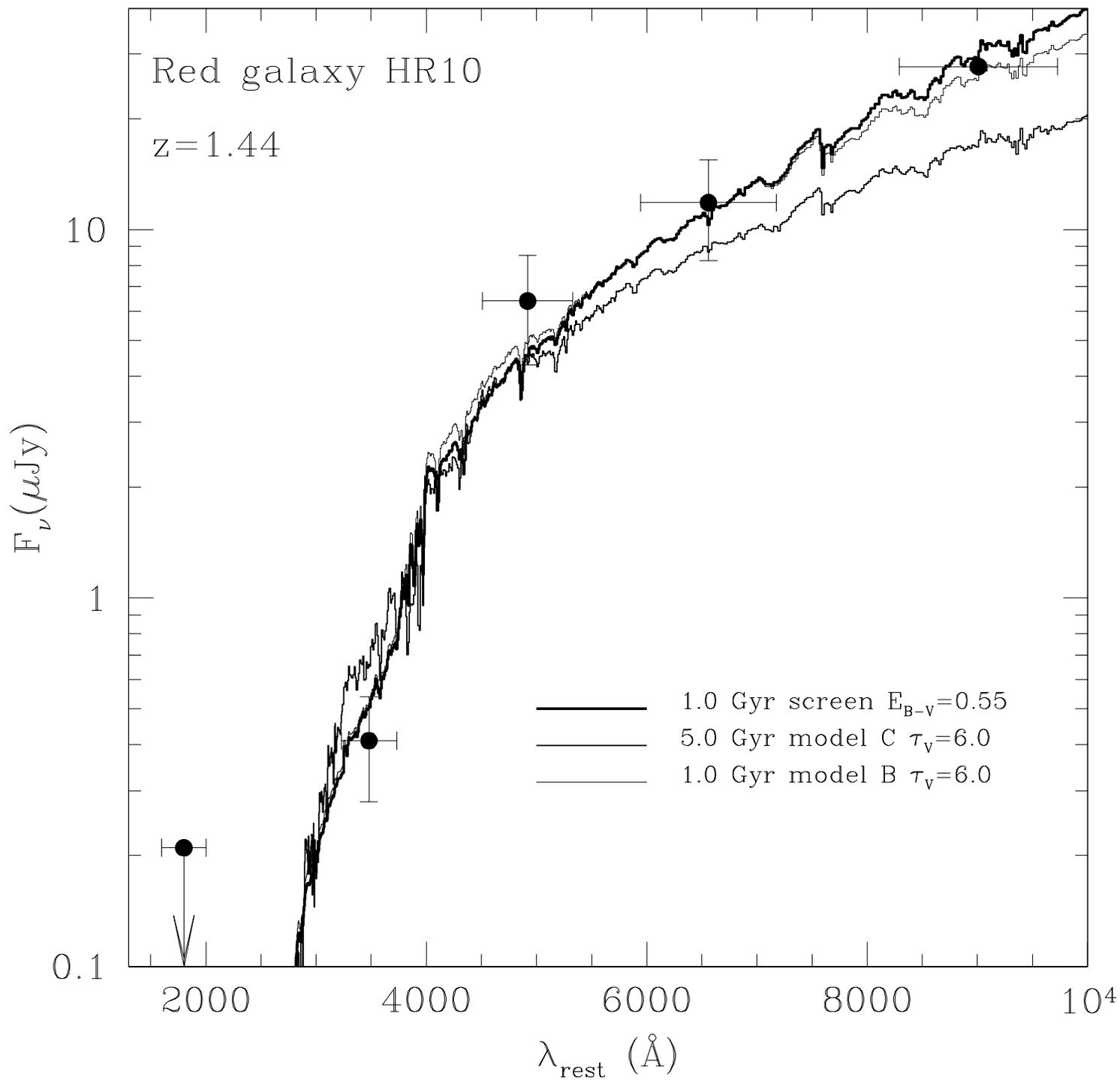
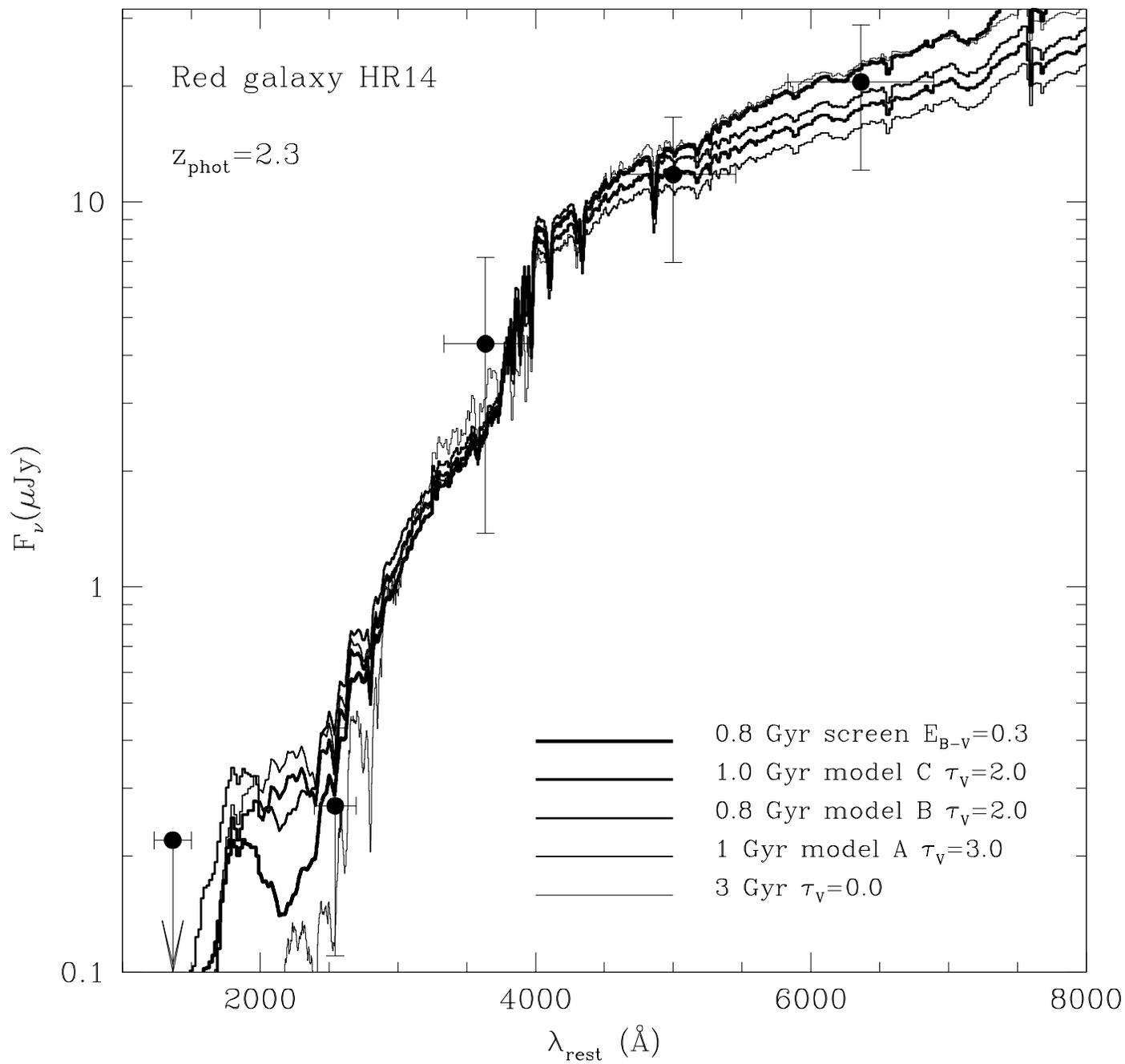
\end{document}